\documentclass[
final
  ]
  {aipproc}

\layoutstyle{6x9}

\begin{document}

\title 
      [ GRAVITATIONAL WAVES AND  GRB]
      {Gravitational Waves and GRBs from Tidal Disruption of Stars
in the Center of Galaxies}


\author{P. Fortini}{
  address={Department of Physics, University of Ferrara and INFN Sezione di Ferrara, via Paradiso 12, 44100 Ferrara, Italy}
}

\author{A. Ortolan}{
  address={INFN - National Laboratories of Legnaro, viale dell'Universit\`a 2, 35020 Legnaro (PD), Italy}
}



\begin{abstract}
Recent measurements by the Chandra satellite have shown that a supermassive 
black hole  of
$M = 2.6 \times 10^{6} M_{\odot}$ is located in the Galactic Center; it seems
probable from other observations that
this fact is common in the majority of galaxies. 
On the other
hand, GRB explosions are typical phenomena linked to galactic
dynamics. In the present paper we discuss the possibility that GRBs are tidal
disruption of stars by supermassive black holes located in the center of 
galaxies. This conjecture can be tested by a gravitational 
wave detector of the class of AURIGA. 

\end{abstract}


\maketitle

\section{Introduction}

GRB engines are characterized by the following facts: 1) they are
point-like sources of electromagnetic energy ($10^{51\div 54}\ erg$) 
comparable to that of a
single galaxy; 2) they are always associated with a host galaxy (see
ref. \cite{gal}). A plausible explanation can be found if there is a 
Supermassive Black Holes (SBH) in the Galactic Center with a mass of 
about $2\times 10^6 \ M_\odot$  
as it was recently discovered by satellite Chandra (se
e \url{http://science.nasa.gov/headlines/y2000/ast29feb_1m.htm} of
Feb. 29, 2000). Moreover, it is very likely that the globular clusters
have in their centers black holes with smaller masses.

The detailed study of the dynamics of the Galactic Center (see  ref.
\cite{SBH}) points out that near the horizon of the SBH all the
galactic objects are crushed by tidal disruption  so that they can
emit both electromagnetic and gravitational waves. For instance
according to this view a GRB can be considered as the end of a star 
while it is swallowed by the SBH.

If this conjecture is correct, it should be possible to detect
gravitational radiation  by means of gravitational wave (gw) detectors using
their directional sensitivity (antenna pattern) towards the candidate sources --
either the Galactic Center or the center of nearby galaxies (M31, M84, M87, 
etc.) or the globular clusters.
The basic idea is to separate the collected events recorded by a 
gravitational wave detector into the {\it ``on source''}
and {\it``off source''} sets depending on the sidereal hours when the 
detector is pointing toward a given source. 
The ``off source'' events will give
an estimate of the background events (from unmodeled noise sources) while
gw bursts can be detected as a mean excess of event energy in the ``on source'' set.

The paper is divided into two parts: in Section 2 we  discuss some
theoretical problems connected with the SBH model and in Section 3 we
describe a method for the measurement of gw emission from
the Galactic Center by means of one or more gw detectors.

\section{Theoretical Problems of SBH}

The paper by Ayal et al. \cite{ayal} is very important because,
in our opinion, it was the first attempt to calculate,
using the 
1PN approximation of General Relativity, what happens to a star falling
into the SBH existing in the center of the Galaxy.
However, these calculations are in some ways too rough, in particular:

\begin{enumerate}
\item
An object like the sun is modeled as a
polytropic fluid in a spherical configuration: only in this way it can be 
treated mathematically, but its 
physical structure is completely ignored. 

\item
Near a few Schwarzschild radii the classical mechanics break down and the
system of a star and a black hole, which has well known obvious closed form solutions,
has no relativistic closed form solution 

\item
The various approximations in general don't converge and the usual result
is the appearance of divergences which make the two body problem
mathematically intractable.  These issues are
covered in Chapter 9 of ref. \cite{SBH}. Here we quote the
main results. The singularities are of three kinds: i) a singularity in the
density: stars cannot exist nearer than the horizon, and effectively they are
destroyed before that point, either by collisions or by the SMB
tidal field. In ref. \cite{ayal} it is shown with explicit calculation
that a polytropic sphere similar to the sun is destroyed at a distance of
approximately $ 15 R_{Sch}$ where
$R_{Sch}$ is the SMB 
horizon; ii) a singularity in
the velocity which makes the dynamics mathematically intractable because of the
Keplerian velocity divergency; iii) an optical
singularity due to
the fact that any mass bends light and, close to massive black hole and
in small
regions, some caustics appear, which cause the kinematics to be equally
divergent.  

\item
Due to the severe limitation of the singularity of the calculations it is very
difficult to believe the numbers put forth by Ayal et al. \cite{ayal}.
For instance, the energy emitted in gw is
$ \sim 1.6 \cdot 10^{46} erg$ and it could be underestimated by orders
of magnitude. This
value ensures that, on one hand the 1PN approximation holds 
and on the other the gravitational interaction is sufficiently strong
to destroy the star.   
It should be noted that the energy converted in gw is calculated by the usual
quadrupole formula and therefore, if it were possible to push the 
approximation 
closer to the black hole horizon, one would expect a greater gw emission. 
\end{enumerate}

Having made the above points, it would be highly desirable to pass
directly to the measurement of gws, in order to test the various    
approximations and the effects of non-linearities 
in the mechanism of gw emission.

\section{Description of the Measurement}
The gravitational wave detector AURIGA, which is operating at the National
Laboratories of Legnaro (Italy), is essentially an aluminum cylinder equipped 
with
a resonant capacitive transducer which is coupled to a high sensitivity
dc-squid amplifier \cite{jpz}. 
\begin{figure}[htbp]
\caption{Relative sensitivity of the AURIGA detector towards the Galactic
Center.}
\includegraphics[height=.22\textheight]{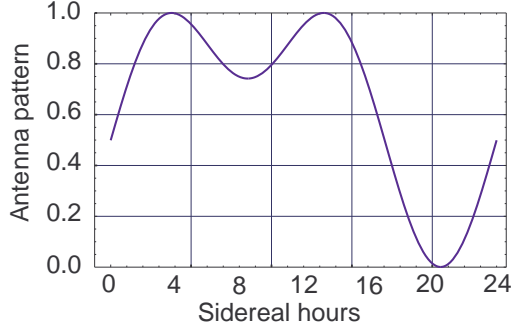}
\label{fig0}
\end{figure}
Due to the combination of
the Earth's rotation and the antenna pattern, two times a day AURIGA
is much more sensitive to the gw flux from the Galactic Center (see Figure 
~\ref{fig0}).
This particular combination happens at the same sidereal hours, when the 
Galactic Center
direction is orthogonal to the detector symmetry axis. The modulation of the gw
flux
must have a period of one sidereal day with the two characteristic maxima of Figure~\ref{fig0} occurring 
at sidereal hours 4 and 13. This peculiar signature of the gw signals can be
used to test if the conjecture we presented in
the preceding Sections is true.
In fact, the Galactic Center is the place
where a high flux of gravitational waves should be emitted by stars
and gas swallowed by the black hole. The cumulative counts of detected events 
should exhibit two peaks separated by 11 hours.
If we collect data for a
sufficiently long time ($\sim 100$ sidereal days),  
 we can form two distinct set of
events: the ``on source'' set which collects all the
events occurring around sidereal hours 4 and 13, and the ``off source'' sets
corresponding to events around sidereal hour 20. The ``off source'' events 
will give 
an estimate of the background events (from unmodeled noise sources)
while gravitational wave bursts (GWBs)
can be detected as an excess of event energy (in average) for events 
belonging to the ``on source'' set.

The Mann--Whitney test, also known as the rank sum test or U--test \cite{mw}, can be used to 
falsify  
the ``null hypothesis'' that the two populations ``on'' and ``off'' are identical, i.e. 
that the gw candidate events (belonging to the ``on source'' set) have the same mean energy $\langle E \rangle $
of background events (belonging to the ``off source set''). The mean energy captured by the 
detector for the two sets can be expressed as  
\begin{equation}
\langle E_{on}\rangle = \langle E_{off}\rangle +  h^2_{RMS} \ ,
\end{equation}
where  $h^2_{RMS}$ is the averaged GWB amplitude associated with the activity of the Galactic Center. 
The sensitivity of the search depends on three main factors \cite{grb}, namely: i) the minimal GWB amplitude 
detectable at unit signal-to-noise ratio  
$h_{min}\equiv(\tau_s \ (\int_{-\infty}^{+\infty} 1/S_{h}(\nu)d\nu)^{-1/2}$, where $\tau_s$ is the duration of the GWB $(\sim 1\ msec)$ and $S_h(\nu)$ is the power spectral density of the 
 noise expressed in terms of
gravitational wave amplitude at the detector input; ii) the event search threshold, 
usually set to $5\times h_{min}$  and iii) the number of the events in the 
``on'' and ``off'' sets, $N_{on}$ and $N_{off}$ respectively.

\begin{figure}[h]
\caption{Power spectral density of the AURIGA noise expressed in terms of
gravitational wave amplitude at the detector input.}
\includegraphics[height=.3\textheight]{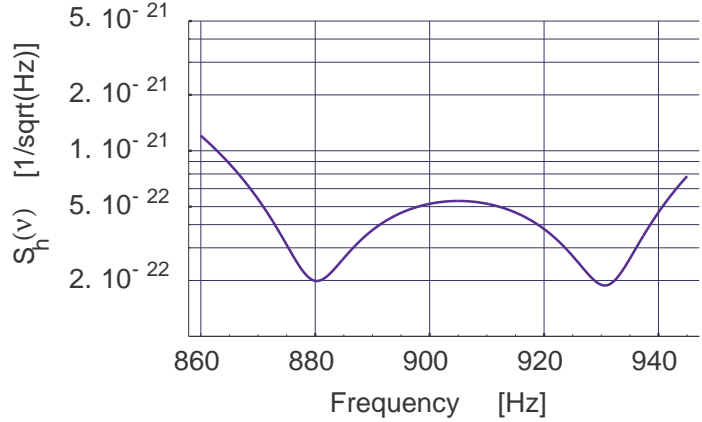}
\label{fig1}
\end{figure}
A similar technique has been applied by our group (see ref. \cite{grb} and 
\cite{grb1}) to test
if there exists a concomitant emission of GRBs and gw bursts. Unfortunately, 
as most GRBs are
at cosmological distances, the AURIGA sensitivity 
($h_{min} \sim 10^{-19}$) and duty cycle ($N_{on}\approx 100$) were not sufficient to detect any
positive effect.
However, in the present paper we limit our analysis at the distance of
the Galactic Center (i.e. $\sim 8\ kpc$), and so the  probability to detect
GWBs is higher. 

In addition,  the new experimental setup of
AURIGA \cite{jpz} promises an increase of its sensitivity  and  bandwidth 
as reported in Figure~\ref{fig1}, which translates into a burst
sensitivity of $h_{min} \sim 10^{-20}$. The same procedure can be applied to 
the events in  coincidence among two or more parallel detectors.
To conclude we must recall that the group of Rome have already tried
to  
detect  
GWBs coming from the galactic plane (see ref. \cite{astone}
and the subsequent criticism in ref. \cite{finn}).


\bibliographystyle{aipproc}
\bibliography{biblio}

\end{document}